\documentstyle[12pt,psfig,axodraw]{article}

\newcommand{\gsim}{\lower.7ex\hbox{$\;\stackrel{\textstyle>}{\sim}\;$}}
\newcommand{\lsim}{\lower.7ex\hbox{$\;\stackrel{\textstyle<}{\sim}\;$}}

\newcommand{\met}{\rlap{\,/}E_T}
\newcommand\beq{\begin{equation}}
\newcommand\eeq{\end{equation}}
\newcommand\bear{\begin{eqnarray}}
\newcommand\eear{\end{eqnarray}}
\def\ap  #1 #2 #3 #4 {Ann.~Phys.         {\bf  #1}, #2 (#3)#4 }
\def\aplb#1 #2 #3 #4 {Acta Phys.~Pol.    {\bf B#1}, #2 (#3)#4 }
\def\cpc #1 #2 #3 #4 {Comp.~Phys.~Comm.  {\bf  #1}, #2 (#3)#4 }
\def\jetp#1 #2 #3 #4 {JETP Lett.         {\bf  #1}, #2 (#3)#4 }
\def\npb #1 #2 #3 #4 {Nucl.~Phys.        {\bf B#1}, #2 (#3)#4 }
\def\mpla#1 #2 #3 #4 {Mod.~Phys.~Lett.   {\bf A#1}, #2 (#3)#4 }
\def\plb #1 #2 #3 #4 {Phys.~Lett.        {\bf B#1}, #2 (#3)#4 }
\def\pr  #1 #2 #3 #4 {Phys.~Rep.         {\bf  #1}, #2 (#3)#4 }
\def\prd #1 #2 #3 #4 {Phys.~Rev.         {\bf D#1}, #2 (#3)#4 }
\def\prl #1 #2 #3 #4 {Phys.~Rev.~Lett.   {\bf  #1}, #2 (#3)#4 }
\def\ptp #1 #2 #3 #4 {Prog.~Theor.~Phys. {\bf  #1}, #2 (#3)#4 }
\def\zpc #1 #2 #3 #4 {Zeit.~Phys.        {\bf C#1}, #2 (#3)#4 }

\begin{document}

\baselineskip=24pt

\thispagestyle{empty}
\begin{titlepage}
\title {\vspace{-2.0cm} 
\hfill {\normalsize FERMILAB--PUB--99/209--T}\\
\vspace{0.9cm} 
New backgrounds in Trilepton, Dilepton and \\
Dilepton plus Tau Jet SUSY Signals at the Tevatron }

\vspace{3cm}

\author{\\
{\sc Konstantin T. Matchev}\\
{\small Theoretical Physics Department}\\
{\small Fermi National Accelerator Laboratory}\\
{\small Batavia, IL 60510}\\
\\
{\sc Damien M. Pierce} \\ 
{\small Physics Department} \\
{\small Brookhaven National Laboratory} \\
{\small Upton, NY 11973}\\
}

\date{}
\maketitle
\thispagestyle{empty}

\begin{abstract}
\noindent
We determine the Tevatron's reach in supersymmetric parameter space in
trilepton, like-sign dilepton, and dilepton plus tau-jet channels,
taking all relevant backgrounds into account. We show results for the
minimal supergravity model. With a standard set of cuts we find that
the previously unaccounted for $W\gamma^\ast$ background is larger
than all other backgrounds combined. We include cuts on the dilepton
invariant mass and the $W$-boson transverse mass to reduce the
$W\gamma^\ast$ background to a reasonable level.  We optimize cuts at
each point in supersymmetry parameter space in order to maximize
signal-to-noise.

\end{abstract}

\vspace{1.8cm}
\centerline{\sl Submitted to Phys. Lett. B}

\end{titlepage}

\setcounter{page}{1} 

\section{Introduction} \label{sec:introduction}

Low-energy supersymmetry (SUSY) is the most popular extension of the
Standard Model (SM). It is vigorously sought at LEP, and has been and
will continue to be actively looked for in the previous and
forthcoming runs of the Fermilab Tevatron Collider~\cite{Workshop
Report}.

There are many possible manifestations of low-energy SUSY.  With more
than 100 new parameters, theorists have out of necessity invented high
scale models with drastically fewer parameters. These models can have
qualitatively distinct low-energy spectra, leading to a variety of
collider signatures. In this paper we explore the reach of the
Tevatron in the parameter space of the most commonly considered model,
the minimal supergravity model~\cite{msugra}. As in many models, this
model respects gaugino mass unification. This implies that in the
physical low-energy spectrum the electroweak gauginos are
significantly lighter than the gluino, so that their production cross
sections are the largest in the allowed regions of parameter space. In
addition to a large production cross section we want a significant
branching fraction into a channel with relatively small Standard Model
background. With all this in mind, the trilepton (3L) signal,
$\ell^\pm\ell^+\ell^-\met$ with $\ell=e$ or $\mu$, has been considered
a gold-plated mode for SUSY discovery at the Tevatron \cite{3L -
ancient, 3L, Mrenna, BK, LM, MP-3L, BDPQT}, prompting several Run I
analyses at the Tevatron~\cite{3L-exp}. The 3L signal is mainly
produced via $p\bar p \rightarrow \tilde\chi_2^0\tilde\chi_1^+$. Being
one of the most extensively studied channels for SUSY discovery, it
was naturally among the main focal points of the Run II Workshop
\cite{Workshop Report}, where the emphasis was placed on optimizing
the analysis cuts in order to maximize the Run II Tevatron reach.

In a recent paper \cite{MP-3L}, we took this approach further by
considering thousands of sets of cuts in order to determine which one
gives the best reach. We also supplemented our trilepton SUSY search
with two other promising signatures --- the inclusive like-sign
dilepton~\cite{JN} and `dilepton plus a tau jet'~\cite{LM}
channels.  Along with the mandatory plots of the Tevatron reach in
parameter space, the main result from \cite{MP-3L} was that the SM
background has been grossly underestimated in the previous studies
(\cite{3L,BK}, and to some extent in \cite{Mrenna}).  We traced the
main cause of the problem to the inadequacy of the event generator
ISAJET \cite{ISAJET} in simulating the SM $WZ$ and $ZZ$
backgrounds. In ISAJET the zero width approximation is used in
generating both $WZ$ and $ZZ$. In PYTHIA \cite{PYTHIA}, a Breit-Wigner
distribution is used for the $W$- and $Z$-bosons. The finite $Z$-width
leads to broader dilepton spectra and hence significantly larger
background.

The $W\gamma^\ast$ background is not incorporated in either ISAJET or
PYTHIA. Hence, it has not been taken into account in previous studies
(see, however, Refs.~\cite{CE,BDPQT}). We find that this background is
larger than all previously considered backgrounds combined. In light
of the importance of the trilepton channel for Run II, we are
compelled to update our analysis of Ref.~\cite{MP-3L}.

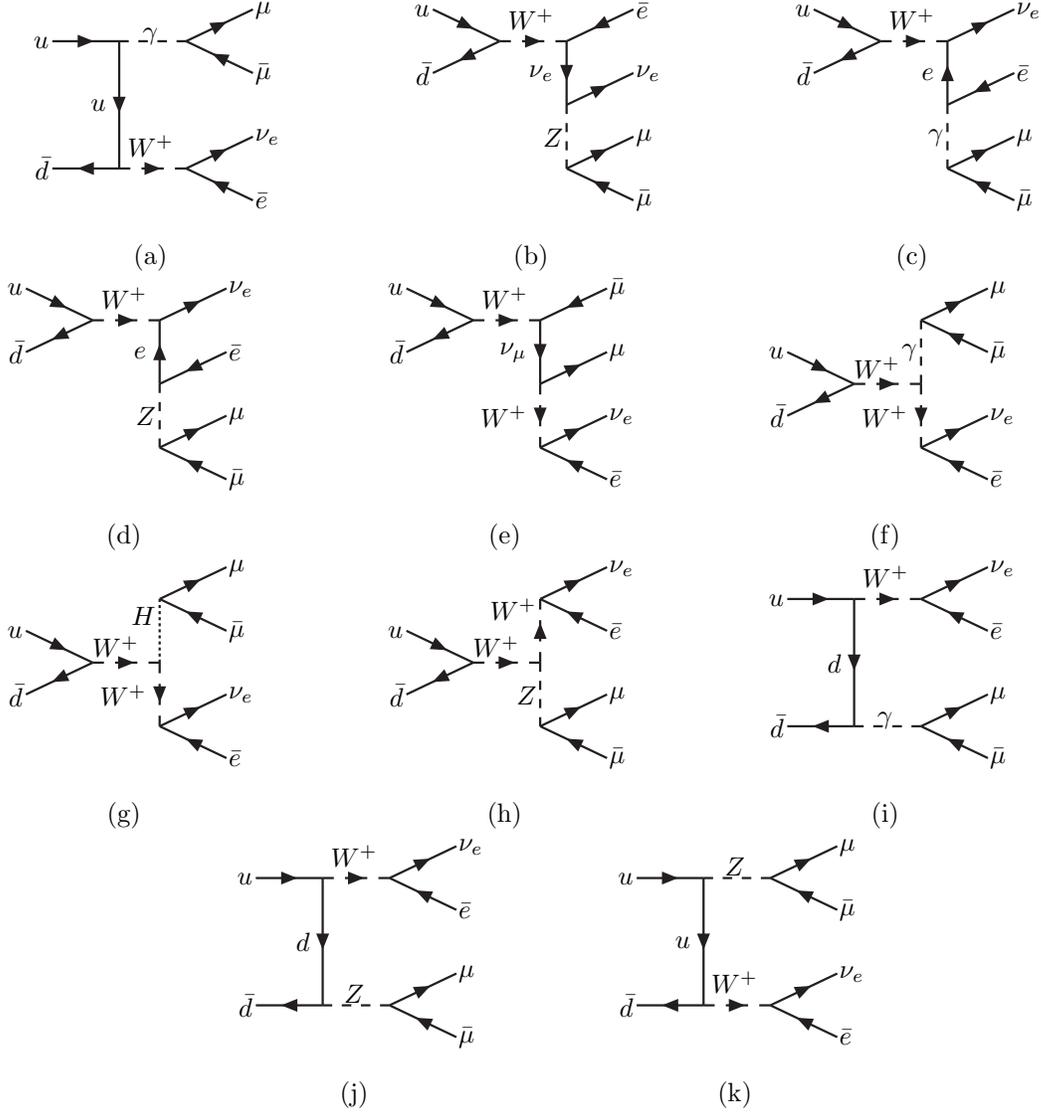
\begin{figure}[t!]
\begin{center}
{
\unitlength=1.2 pt
\SetScale{1.2}
\SetWidth{0.7}      
\footnotesize         
{} \qquad\allowbreak
\begin{picture}(95,99)(0,0)
\Text(15.0,80.0)[r]{$u$}
\ArrowLine(16.0,80.0)(37.0,80.0) 
\Text(47.0,81.0)[b]{$\gamma$}
\DashLine(37.0,80.0)(58.0,80.0){3.0} 
\Text(80.0,90.0)[l]{$\mu$}
\ArrowLine(58.0,80.0)(79.0,90.0) 
\Text(80.0,70.0)[l]{$\bar{\mu}$}
\ArrowLine(79.0,70.0)(58.0,80.0) 
\Text(33.0,60.0)[r]{$u$}
\ArrowLine(37.0,80.0)(37.0,40.0) 
\Text(15.0,40.0)[r]{$\bar{d}$}
\ArrowLine(37.0,40.0)(16.0,40.0) 
\Text(47.0,44.0)[b]{$W^+$}
\DashArrowLine(37.0,40.0)(58.0,40.0){3.0} 
\Text(80.0,50.0)[l]{$\nu_e$}
\ArrowLine(58.0,40.0)(79.0,50.0) 
\Text(80.0,30.0)[l]{$\bar{e}$}
\ArrowLine(79.0,30.0)(58.0,40.0) 
\Text(47,10)[b] {(a)}
\end{picture} \ 
{} \qquad\allowbreak
\begin{picture}(95,99)(0,0)
\Text(15.0,90.0)[r]{$u$}
\ArrowLine(16.0,90.0)(37.0,80.0) 
\Text(15.0,70.0)[r]{$\bar{d}$}
\ArrowLine(37.0,80.0)(16.0,70.0) 
\Text(47.0,84.0)[b]{$W^+$}
\DashArrowLine(37.0,80.0)(58.0,80.0){3.0} 
\Text(80.0,90.0)[l]{$\bar{e}$}
\ArrowLine(79.0,90.0)(58.0,80.0) 
\Text(54.0,70.0)[r]{$\nu_e$}
\ArrowLine(58.0,80.0)(58.0,60.0) 
\Text(80.0,70.0)[l]{$\nu_e$}
\ArrowLine(58.0,60.0)(79.0,70.0) 
\Text(57.0,50.0)[r]{$Z$}
\DashLine(58.0,60.0)(58.0,40.0){3.0} 
\Text(80.0,50.0)[l]{$\mu$}
\ArrowLine(58.0,40.0)(79.0,50.0) 
\Text(80.0,30.0)[l]{$\bar{\mu}$}
\ArrowLine(79.0,30.0)(58.0,40.0) 
\Text(47,10)[b] {(b)}
\end{picture} \ 
{} \qquad\allowbreak
\begin{picture}(95,99)(0,0)
\Text(15.0,90.0)[r]{$u$}
\ArrowLine(16.0,90.0)(37.0,80.0) 
\Text(15.0,70.0)[r]{$\bar{d}$}
\ArrowLine(37.0,80.0)(16.0,70.0) 
\Text(47.0,84.0)[b]{$W^+$}
\DashArrowLine(37.0,80.0)(58.0,80.0){3.0} 
\Text(80.0,90.0)[l]{$\nu_e$}
\ArrowLine(58.0,80.0)(79.0,90.0) 
\Text(54.0,70.0)[r]{$e$}
\ArrowLine(58.0,60.0)(58.0,80.0) 
\Text(80.0,70.0)[l]{$\bar{e}$}
\ArrowLine(79.0,70.0)(58.0,60.0) 
\Text(57.0,50.0)[r]{$\gamma$}
\DashLine(58.0,60.0)(58.0,40.0){3.0} 
\Text(80.0,50.0)[l]{$\mu$}
\ArrowLine(58.0,40.0)(79.0,50.0) 
\Text(80.0,30.0)[l]{$\bar{\mu}$}
\ArrowLine(79.0,30.0)(58.0,40.0) 
\Text(47,10)[b] {(c)}
\end{picture} \ 
{} \qquad\allowbreak
\\
\vspace*{-0.5cm}
\begin{picture}(95,99)(0,0)
\Text(15.0,90.0)[r]{$u$}
\ArrowLine(16.0,90.0)(37.0,80.0) 
\Text(15.0,70.0)[r]{$\bar{d}$}
\ArrowLine(37.0,80.0)(16.0,70.0) 
\Text(47.0,84.0)[b]{$W^+$}
\DashArrowLine(37.0,80.0)(58.0,80.0){3.0} 
\Text(80.0,90.0)[l]{$\nu_e$}
\ArrowLine(58.0,80.0)(79.0,90.0) 
\Text(54.0,70.0)[r]{$e$}
\ArrowLine(58.0,60.0)(58.0,80.0) 
\Text(80.0,70.0)[l]{$\bar{e}$}
\ArrowLine(79.0,70.0)(58.0,60.0) 
\Text(57.0,50.0)[r]{$Z$}
\DashLine(58.0,60.0)(58.0,40.0){3.0} 
\Text(80.0,50.0)[l]{$\mu$}
\ArrowLine(58.0,40.0)(79.0,50.0) 
\Text(80.0,30.0)[l]{$\bar{\mu}$}
\ArrowLine(79.0,30.0)(58.0,40.0) 
\Text(47,10)[b] {(d)}
\end{picture} \ 
{} \qquad\allowbreak
\begin{picture}(95,99)(0,0)
\Text(15.0,90.0)[r]{$u$}
\ArrowLine(16.0,90.0)(37.0,80.0) 
\Text(15.0,70.0)[r]{$\bar{d}$}
\ArrowLine(37.0,80.0)(16.0,70.0) 
\Text(47.0,84.0)[b]{$W^+$}
\DashArrowLine(37.0,80.0)(58.0,80.0){3.0} 
\Text(80.0,90.0)[l]{$\bar{\mu}$}
\ArrowLine(79.0,90.0)(58.0,80.0) 
\Text(54.0,70.0)[r]{$\nu_\mu$}
\ArrowLine(58.0,80.0)(58.0,60.0) 
\Text(80.0,70.0)[l]{$\mu$}
\ArrowLine(58.0,60.0)(79.0,70.0) 
\Text(54.0,50.0)[r]{$W^+$}
\DashArrowLine(58.0,60.0)(58.0,40.0){3.0} 
\Text(80.0,50.0)[l]{$\nu_e$}
\ArrowLine(58.0,40.0)(79.0,50.0) 
\Text(80.0,30.0)[l]{$\bar{e}$}
\ArrowLine(79.0,30.0)(58.0,40.0) 
\Text(47,10)[b] {(e)}
\end{picture} \ 
{} \qquad\allowbreak
\begin{picture}(95,99)(0,0)
\Text(15.0,70.0)[r]{$u$}
\ArrowLine(16.0,70.0)(37.0,60.0) 
\Text(15.0,50.0)[r]{$\bar{d}$}
\ArrowLine(37.0,60.0)(16.0,50.0) 
\Text(37.0,62.0)[lb]{$W^+$}
\DashArrowLine(37.0,60.0)(58.0,60.0){3.0} 
\Text(57.0,70.0)[r]{$\gamma$}
\DashLine(58.0,60.0)(58.0,80.0){3.0} 
\Text(80.0,90.0)[l]{$\mu$}
\ArrowLine(58.0,80.0)(79.0,90.0) 
\Text(80.0,70.0)[l]{$\bar{\mu}$}
\ArrowLine(79.0,70.0)(58.0,80.0) 
\Text(54.0,50.0)[r]{$W^+$}
\DashArrowLine(58.0,60.0)(58.0,40.0){3.0} 
\Text(80.0,50.0)[l]{$\nu_e$}
\ArrowLine(58.0,40.0)(79.0,50.0) 
\Text(80.0,30.0)[l]{$\bar{e}$}
\ArrowLine(79.0,30.0)(58.0,40.0) 
\Text(47,10)[b] {(f)}
\end{picture} \ 
{} \qquad\allowbreak
\\
\vspace*{-0.5cm}
\begin{picture}(95,99)(0,0)
\Text(15.0,70.0)[r]{$u$}
\ArrowLine(16.0,70.0)(37.0,60.0) 
\Text(15.0,50.0)[r]{$\bar{d}$}
\ArrowLine(37.0,60.0)(16.0,50.0) 
\Text(37.0,62.0)[lb]{$W^+$}
\DashArrowLine(37.0,60.0)(58.0,60.0){3.0} 
\Text(57.0,75.0)[r]{$H$}
\DashLine(58.0,60.0)(58.0,80.0){1.0}
\Text(80.0,90.0)[l]{$\mu$}
\ArrowLine(58.0,80.0)(79.0,90.0) 
\Text(80.0,70.0)[l]{$\bar{\mu}$}
\ArrowLine(79.0,70.0)(58.0,80.0) 
\Text(54.0,50.0)[r]{$W^+$}
\DashArrowLine(58.0,60.0)(58.0,40.0){3.0} 
\Text(80.0,50.0)[l]{$\nu_e$}
\ArrowLine(58.0,40.0)(79.0,50.0) 
\Text(80.0,30.0)[l]{$\bar{e}$}
\ArrowLine(79.0,30.0)(58.0,40.0) 
\Text(47,10)[b] {(g)}
\end{picture} \ 
{} \qquad\allowbreak
\begin{picture}(95,99)(0,0)
\Text(15.0,70.0)[r]{$u$}
\ArrowLine(16.0,70.0)(37.0,60.0) 
\Text(15.0,50.0)[r]{$\bar{d}$}
\ArrowLine(37.0,60.0)(16.0,50.0) 
\Text(37.0,62.0)[lb]{$W^+$}
\DashArrowLine(37.0,60.0)(58.0,60.0){3.0} 
\Text(57.0,77.0)[r]{$W^+$}
\DashArrowLine(58.0,60.0)(58.0,80.0){3.0} 
\Text(80.0,90.0)[l]{$\nu_e$}
\ArrowLine(58.0,80.0)(79.0,90.0) 
\Text(80.0,70.0)[l]{$\bar{e}$}
\ArrowLine(79.0,70.0)(58.0,80.0) 
\Text(57.0,50.0)[r]{$Z$}
\DashLine(58.0,60.0)(58.0,40.0){3.0} 
\Text(80.0,50.0)[l]{$\mu$}
\ArrowLine(58.0,40.0)(79.0,50.0) 
\Text(80.0,30.0)[l]{$\bar{\mu}$}
\ArrowLine(79.0,30.0)(58.0,40.0) 
\Text(47,10)[b] {(h)}
\end{picture} \ 
{} \qquad\allowbreak
\begin{picture}(95,99)(0,0)
\Text(15.0,80.0)[r]{$u$}
\ArrowLine(16.0,80.0)(37.0,80.0) 
\Text(47.0,84.0)[b]{$W^+$}
\DashArrowLine(37.0,80.0)(58.0,80.0){3.0} 
\Text(80.0,90.0)[l]{$\nu_e$}
\ArrowLine(58.0,80.0)(79.0,90.0) 
\Text(80.0,70.0)[l]{$\bar{e}$}
\ArrowLine(79.0,70.0)(58.0,80.0) 
\Text(33.0,60.0)[r]{$d$}
\ArrowLine(37.0,80.0)(37.0,40.0) 
\Text(15.0,40.0)[r]{$\bar{d}$}
\ArrowLine(37.0,40.0)(16.0,40.0) 
\Text(47.0,41.0)[b]{$\gamma$}
\DashLine(37.0,40.0)(58.0,40.0){3.0} 
\Text(80.0,50.0)[l]{$\mu$}
\ArrowLine(58.0,40.0)(79.0,50.0) 
\Text(80.0,30.0)[l]{$\bar{\mu}$}
\ArrowLine(79.0,30.0)(58.0,40.0) 
\Text(47,10)[b] {(i)}
\end{picture} \ 
{} \qquad\allowbreak
\\
\vspace*{-0.5cm}
\begin{picture}(95,99)(0,0)
\Text(15.0,80.0)[r]{$u$}
\ArrowLine(16.0,80.0)(37.0,80.0) 
\Text(47.0,84.0)[b]{$W^+$}
\DashArrowLine(37.0,80.0)(58.0,80.0){3.0} 
\Text(80.0,90.0)[l]{$\nu_e$}
\ArrowLine(58.0,80.0)(79.0,90.0) 
\Text(80.0,70.0)[l]{$\bar{e}$}
\ArrowLine(79.0,70.0)(58.0,80.0) 
\Text(33.0,60.0)[r]{$d$}
\ArrowLine(37.0,80.0)(37.0,40.0) 
\Text(15.0,40.0)[r]{$\bar{d}$}
\ArrowLine(37.0,40.0)(16.0,40.0) 
\Text(47.0,41.0)[b]{$Z$}
\DashLine(37.0,40.0)(58.0,40.0){3.0} 
\Text(80.0,50.0)[l]{$\mu$}
\ArrowLine(58.0,40.0)(79.0,50.0) 
\Text(80.0,30.0)[l]{$\bar{\mu}$}
\ArrowLine(79.0,30.0)(58.0,40.0) 
\Text(47,10)[b] {(j)}
\end{picture} \ 
{} \qquad\allowbreak
\begin{picture}(95,99)(0,0)
\Text(15.0,80.0)[r]{$u$}
\ArrowLine(16.0,80.0)(37.0,80.0) 
\Text(47.0,81.0)[b]{$Z$}
\DashLine(37.0,80.0)(58.0,80.0){3.0} 
\Text(80.0,90.0)[l]{$\mu$}
\ArrowLine(58.0,80.0)(79.0,90.0) 
\Text(80.0,70.0)[l]{$\bar{\mu}$}
\ArrowLine(79.0,70.0)(58.0,80.0) 
\Text(33.0,60.0)[r]{$u$}
\ArrowLine(37.0,80.0)(37.0,40.0) 
\Text(15.0,40.0)[r]{$\bar{d}$}
\ArrowLine(37.0,40.0)(16.0,40.0) 
\Text(47.0,44.0)[b]{$W^+$}
\DashArrowLine(37.0,40.0)(58.0,40.0){3.0} 
\Text(80.0,50.0)[l]{$\nu_e$}
\ArrowLine(58.0,40.0)(79.0,50.0) 
\Text(80.0,30.0)[l]{$\bar{e}$}
\ArrowLine(79.0,30.0)(58.0,40.0) 
\Text(47,10)[b] {(k)}
\end{picture} \ 
}
\parbox{5.5in}{
\caption[] {\small The diagrams for the
$p\bar{p}\rightarrow W^+(Z/\gamma^\ast)\rightarrow 
{\ell'}^+\bar{\nu}_{\ell'}\ell^+\ell^-$ background.
Here $u$ and $d$ stand for a generic up-type and down-type quark,
respectively. 
\label{diagrams}}}
\end{center}
\end{figure}

We were faced with several options as to how to incorporate the
$W\gamma^\ast$ process\footnote{By $W\gamma^\ast$ we implicitly refer
to the $Z$-$\gamma^\ast$ interference as well.}. There are several
parton level Monte Carlo generators which use the full set of diagrams
(see Fig.~\ref{diagrams}) to generate what is loosely called ``$WZ$'',
but in reality is the $2\rightarrow4$ process $p\bar{p}\rightarrow
\ell^\pm\nu\ell'^+\ell'^-$. Three such generators are MADGRAPH
\cite{madgraph}, COMPHEP \cite{comphep} and MCFM \cite{CE}.
The choice of a particular generator is dictated by a matter of
convenience and/or experience. We want to not only generate $WZ$
events with the correct kinematics, but also to include a full
detector simulation as we did in \cite{MP-3L}, making use of the SHW
package \cite{SHW,TAUOLA,STDHEP}.  In addition, to make the simulation
fully realistic, we need to include the effects from initial and final
state radiation (ISR,FSR), therefore we cannot just link one of the
leading order parton level Monte Carlos to our detector simulation
package.  What we choose to do instead is to use COMPHEP to generate
hard scattering events at leading order, then we pipe those through
PYTHIA which adds showering and hadronization, and finally we run the
result through SHW
\footnote{Alternatively, one can omit the first step and generate the
$WZ$ events directly from PYTHIA, reweighting the events so as to fit
the distributions of a few key variables (e.g. dilepton invariant
mass, lepton $p_T$ spectrum or angular distributions, etc.).  In the
early stages of this project we followed this approach and reweighted
the PYTHIA events to fit the invariant mass distribution from MCFM. We
then applied the same cut optimization procedure as in \cite{MP-3L},
and presented our results for the Tevatron reach in a series of talks
\cite{talks}. This procedure is, of course, only an attempt to
approximate what we are doing here. The results turn out to be in
reasonable agreement with the current results.}. The resulting
parton-level cross section was integrated with the {\tt CTEQ4m}
structure functions \cite{CTEQ}.

Unfortunately, with a standard set of cuts \cite{BK} the
$W\gamma^\ast$ background is about 2.7 fb, which is larger than all
previously considered backgrounds combined (2.1 fb~\cite{MP-3L}). This
new source of background dwarfs previous estimates. For example, it is
over 4 times the {\em total} background found in Ref.~\cite{BK}, and
our total background is now more than 8 times the total background
reported in~\cite{BK}.  

These recent developments necessitate the invention of new cuts,
specifically designed to suppress the off-shell $Z/\gamma$ component
of the background.  One obvious variable to consider is the
invariant mass $m_{\ell^+\ell^-}$ of an opposite sign, same flavor
lepton pair in the event.  The inclusion of the off-shell photon
contribution increases the relative weight of events with low
$m_{\ell^+\ell^-}$. In anticipation of this effect, in
Ref.~\cite{MP-3L} we employed a low invariant mass cut of
$m_{\ell^+\ell^-}>12$ GeV.

\begin{figure}[t]
\centerline{\psfig{figure=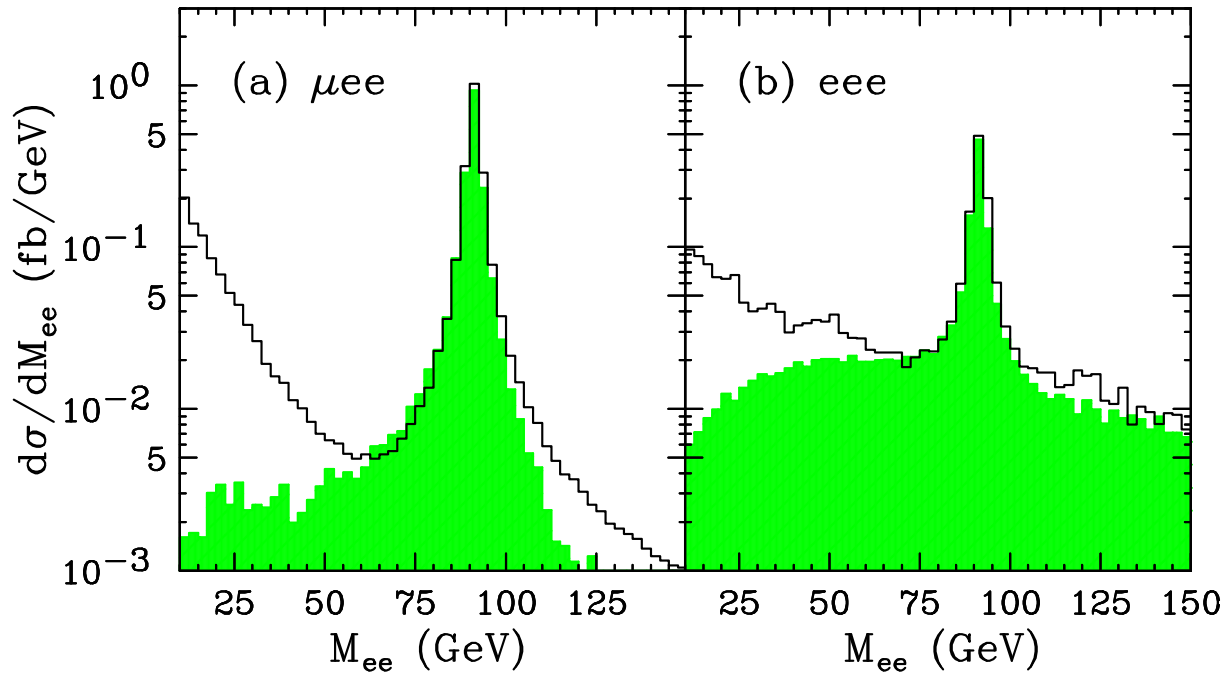,height=3.5in}}
\begin{center}
\parbox{5.5in}{
\caption[] {\small The invariant mass distribution $m_{\ell^+\ell^-}$
of the opposite sign, same flavor leptons in (a) $\mu^\pm e^+e^-$
events and (b) $e^\pm e^+e^-$ events.  The histograms show the results
from COMPHEP and from PYTHIA (shaded). We have imposed
nominal charged lepton cuts $p_T(\ell)>5$ GeV and
$m_{\ell^+\ell^-}>10$ GeV.  Each histogram is normalized to its cross
section.  In (b), we fill both invariant mass combinations, each with
weight $1/2$.
\label{im2l}}}
\end{center}
\end{figure}

In Fig.~\ref{im2l} we show the $m_{\ell^+\ell^-}$ distribution in $WZ$
events from COMPHEP and PYTHIA, before detector simulation and without
ISR/FSR\footnote{In ISAJET the invariant mass distribution of the
leptons is a $\delta$ function at the $Z$-mass.}.
We divide the $WZ$ trilepton sample into opposite flavor (OF)
($e^\pm \mu^+\mu^-$ and $\mu^\pm e^+e^-$) and same flavor (SF)
($\mu^\pm \mu^+\mu^-$ and $e^\pm e^+e^-$) subsets, and show
the results for each subset separately in Fig.~\ref{im2l}a
and Fig.~\ref{im2l}b, respectively. For the OF sample, we
know unambiguously which two leptons came from the
off-shell $Z/\gamma$, so we enter one
invariant mass combination per event. However, in the 
SF sample, there are two possible invariant mass combinations
for each event, 
and there is no way to know which one was from the $Z/\gamma$.
Hence, in Fig.~\ref{im2l}b we enter both combinations, each with
weight $1/2$.

First we see that neglecting the virtual photon contribution and the
$Z-\gamma$ interference leads to a significant underestimate of the
$WZ$ background. In fact, the virtual photon contribution diverges in
the limit $m_{\ell^+\ell^-}\rightarrow 0$!  Second, the low-end
invariant mass cut $m_{\ell^+\ell^-}>12$ GeV that we used in
\cite{MP-3L} is clearly not very efficient in suppressing the
additional $\gamma^\ast$ background and the cut threshold needs to be
increased. The optimum threshold will depend on the signal
distribution, whose shape is controlled by the value of the chargino
mass $m_{\tilde\chi_1^+}$ and is thus parameter space dependent.  We
therefore incorporate the low-end invariant mass cut into our
optimization scheme, and we consider the cuts
$m_{\ell^+\ell^-}^\gamma>\{10$--$60\}$ GeV, in 5 GeV increments.  We
choose the optimal one at each point in SUSY parameter space (for
further details on our optimization procedure, see~\cite{MP-3L}).

\begin{figure}[t]
\centerline{\psfig{figure=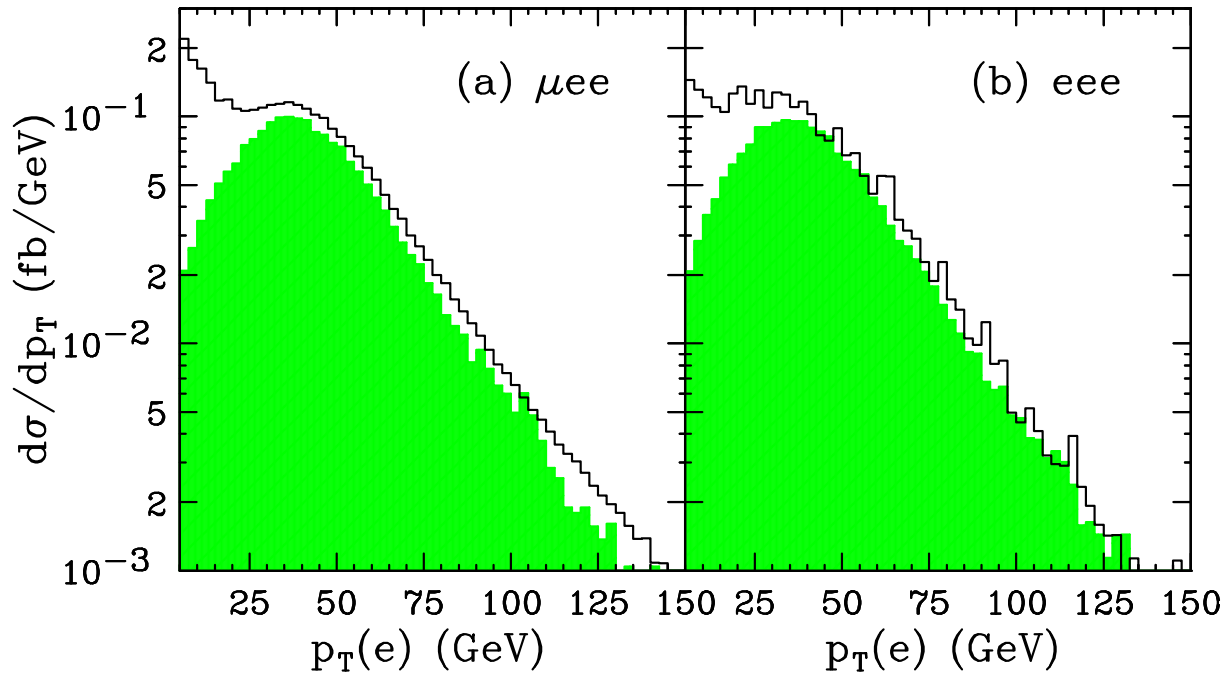,height=3.5in}}
\begin{center}
\parbox{5.5in}{
\caption[] {\small The same as Fig.~\ref{im2l}, but for the
$p_T$ distribution of the leptons possibly coming from the $Z$.
In the case of $\mu^\pm e^+e^-$, we fill the $p_T$ of both
$e^+$ and $e^-$,
each with weight $1/2$. For the case of $e^\pm e^+e^-$
we fill the $p_T$ of the odd-sign lepton with weight $1/2$
and the $p_T$ of the like sign leptons with weight $1/4$ each.
\label{ptl}}}
\end{center}
\end{figure}

In Fig.~\ref{ptl} we compare the COMPHEP and PYTHIA $p_T$ distributions
of the leptons.
We see that most of the additional events due to the $\gamma^\ast$
contribution tend to have small $p_T$.  This implies that the soft
cuts on the lepton $p_T$ introduced in Ref.~\cite{BK} may be
inefficient in removing the new background component. The soft $p_T$
cuts could be detrimental to the reach in regions of parameter space
where the size of the background is important.

Alternatively, Ref.~\cite{BDPQT} suggests a cut on the transverse
mass $m_T$ of any $\ell\nu$ pair which may originate from a $W$-boson.
The advantage of this cut is that it removes background events
irrespective of whether the remaining lepton pair came from a $Z$,
$\gamma^\ast$ or the interference contribution. We shall therefore
optionally incorporate this cut in our analysis of all three channels:
$60<m_T(\ell,\nu)<85$ GeV. The remaining cuts that we use are fully
described in Ref.~\cite{MP-3L} and will not be repeated here.

\begin{figure}[t]
\centerline{\psfig{figure=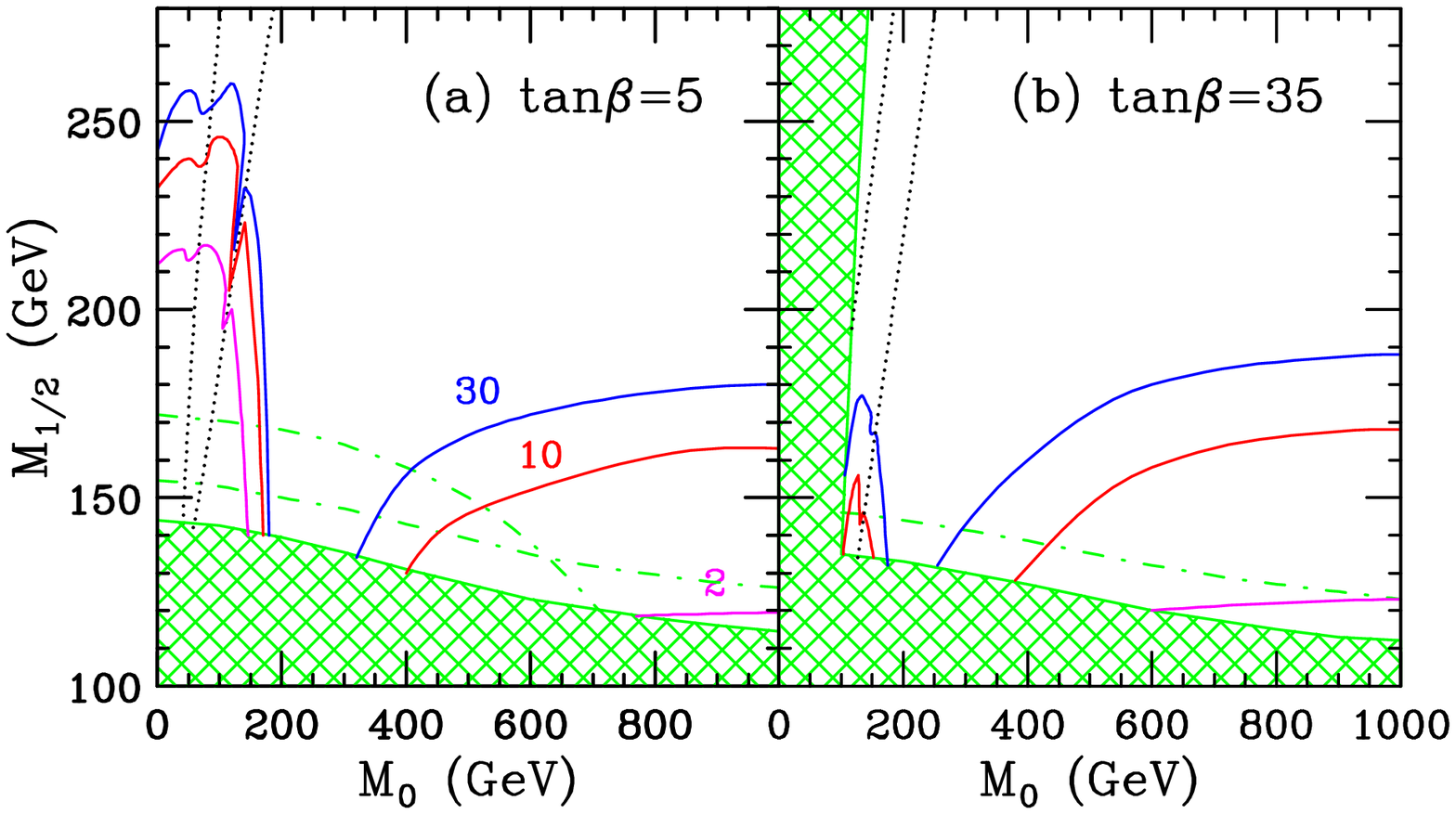,height=3.5in}}
\begin{center}
\parbox{5.5in}{
\caption[] {\small Tevatron reach in the trilepton channel in
the $M_0-M_{1/2}$ plane, for fixed values of $A_0=0$, $\mu>0$
and (a) $\tan\beta=5$, or (b) $\tan\beta=35$. Results are shown
for 2, 10 and 30 ${\rm fb}^{-1}$ total integrated luminosity.
\label{3l}}}
\end{center}
\end{figure}

\begin{figure}[t]
\centerline{\psfig{figure=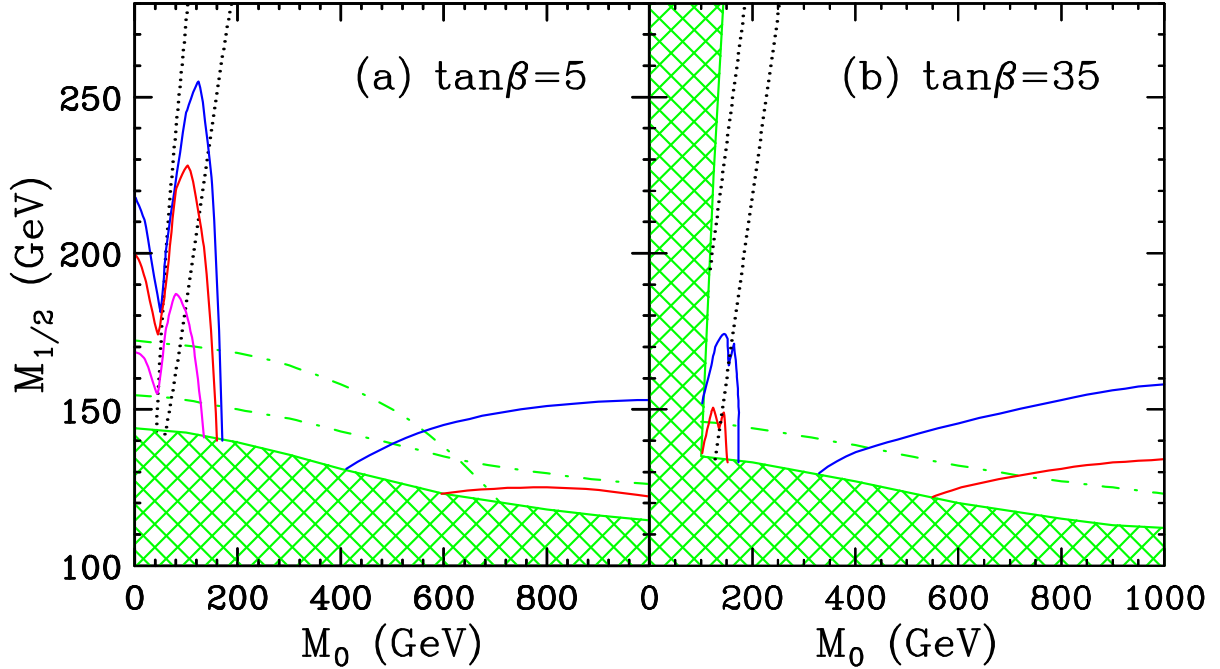,height=3.5in}}
\begin{center}
\parbox{5.5in}{
\caption[] {\small The same as Fig.~\ref{3l}, but for the like-sign
dilepton channel.
\label{2l}}}
\end{center}
\end{figure}

\begin{figure}[t]
\centerline{\psfig{figure=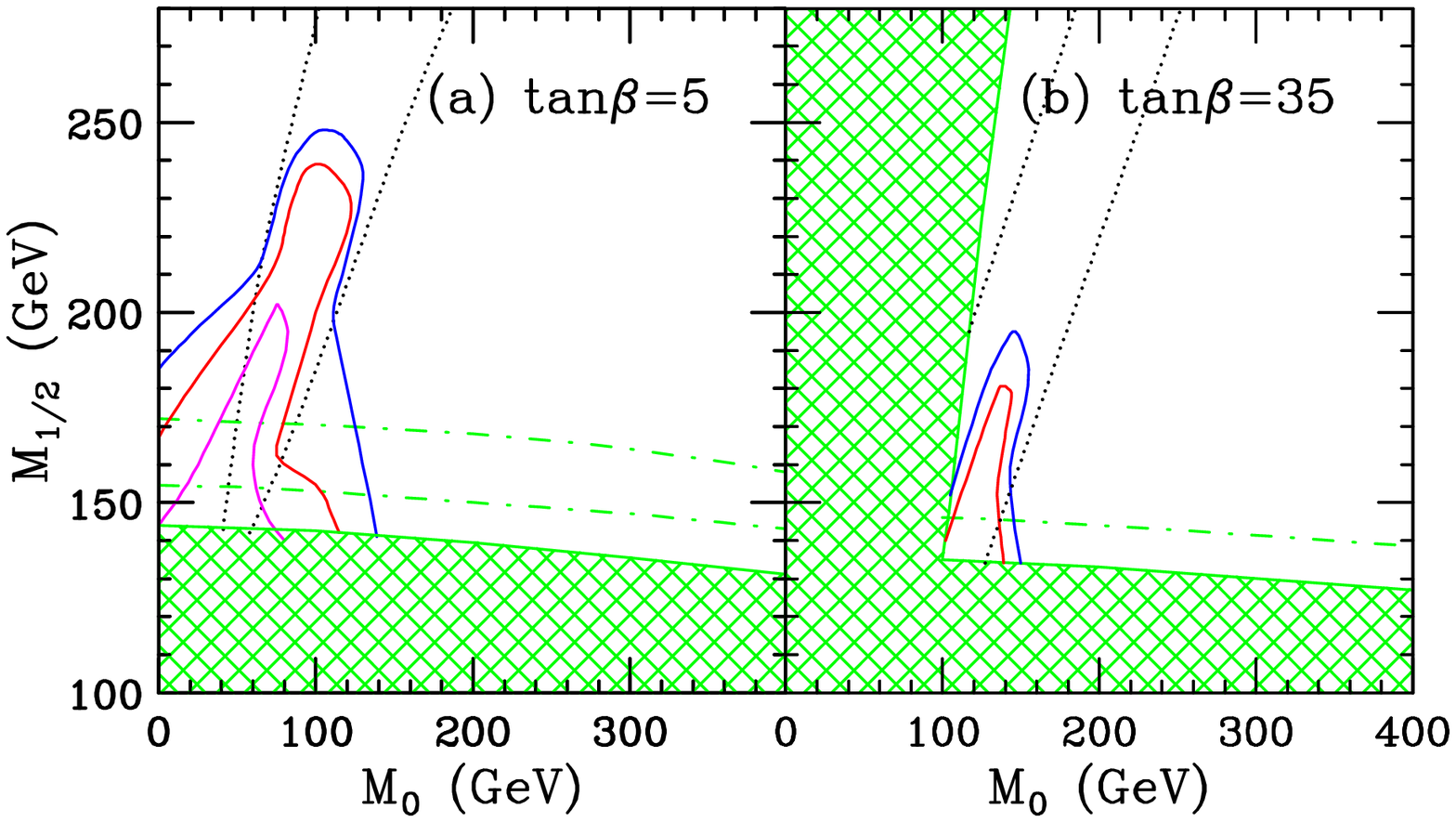,height=3.5in}}
\begin{center}
\parbox{5.5in}{
\caption[] {\small The same as Fig.~\ref{3l}, but for the dilepton
plus a tau jet channel.
\label{2l1t}}}
\end{center}
\end{figure}

We present our results for the Tevatron reach in the trilepton, 
like-sign dilepton and dilepton plus tau jet channels in
Figs.~\ref{3l}, \ref{2l}, and \ref{2l1t}, respectively.
We require the observation of at least 5 signal events, and present
our results as $3\sigma$ exclusion contours in the $M_0-M_{1/2}$
plane, for two representative values of $\tan\beta$, 5 and 35. We fix
$\mu>0$ and $A_0=0$.  The cross-hatched region is excluded by current
limits on the superpartner masses. The dot-dashed lines correspond to
the projected LEP-II reach for the chargino and the lightest Higgs
masses. In Figs.~(a) the left dotted line shows where
$m_{\tilde\nu_\tau}=m_{\tilde\chi_1^\pm}$ and the right dotted line
indicates $m_{\tilde\tau_1}=m_{\tilde\chi_1^\pm}$ (and
$m_{\tilde\tau}\simeq m_{\tilde\mu}\simeq m_{\tilde e}$). In Figs.~(b)
the dotted lines show where $m_{\tilde e_R}=m_{\tilde\chi_1^\pm}$
(left) and $m_{\tilde\tau_1}=m_{\tilde\chi_1^\pm}$ (right).

We see that although the inclusion of the $\gamma^\ast$ background
leads to a significant increase in the raw background cross section,
the reach is somewhat similar to what was presented in
Ref.~\cite{MP-3L}, since the additional cuts help to increase the
signal-to-noise ratio reasonably close to previous levels.  At small
$\tan\beta$ the trilepton channel provides for significant reach at
both small $M_0$ ($M_0\lsim 150$ GeV) and large $M_0$ ($M_0\gsim 400$
GeV). The other channels have somewhat less reach. At large
$\tan\beta$ the dilepton $+~\tau$ jet channel provides the best reach
at small $M_0$ ($M_0\lsim 160$ GeV), while at large $M_0$
($M_0\gsim400$ GeV) the trilepton channel still provides for decent
reach with 30 fb$^{-1}$. With only 2 fb$^{-1}$ the reach is quite
limited.

\begin{figure}[t]
\centerline{\psfig{figure=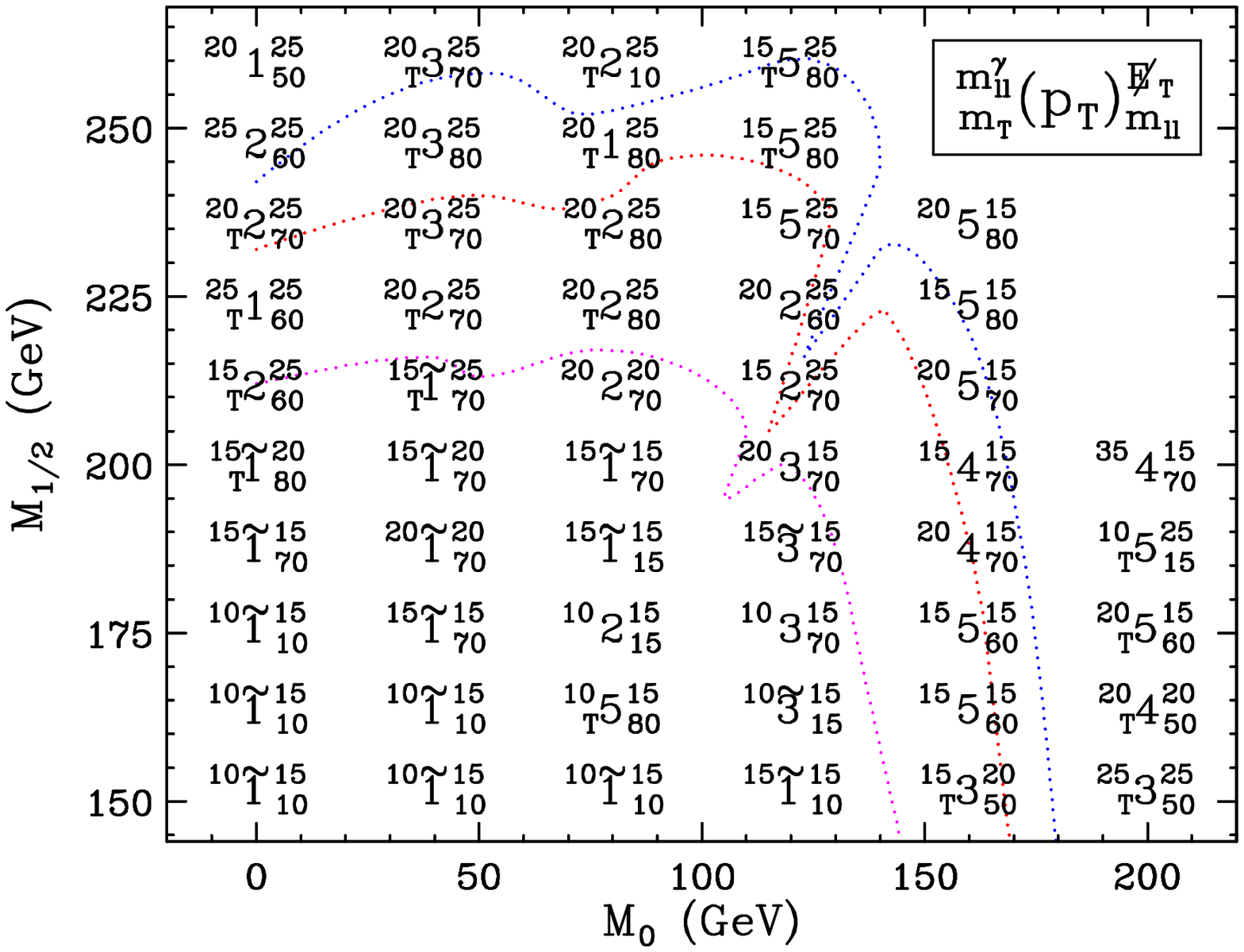,height=3.5in}}
\begin{center}
\parbox{5.5in}{
\caption[] {\small The optimal sets of trilepton cuts in the $M_0,~
M_{1/2}$ plane, for $\tan\beta=5$ and small $M_0$.  We show the
optimal low end dilepton mass cut $m^\gamma_{\ell^+\ell^-}$, missing
$E_T$ cut $\met$, high end dilepton mass cut $m_{\ell^+\ell^-}$,
transverse $\ell\nu$ mass cut and lepton $p_T$ cut (see text). The
dotted lines indicate the reach contours from Fig.~\ref{3l}.
\label{bc5_m0_small}}}
\end{center}
\end{figure}

\begin{figure}[t]
\centerline{\psfig{figure=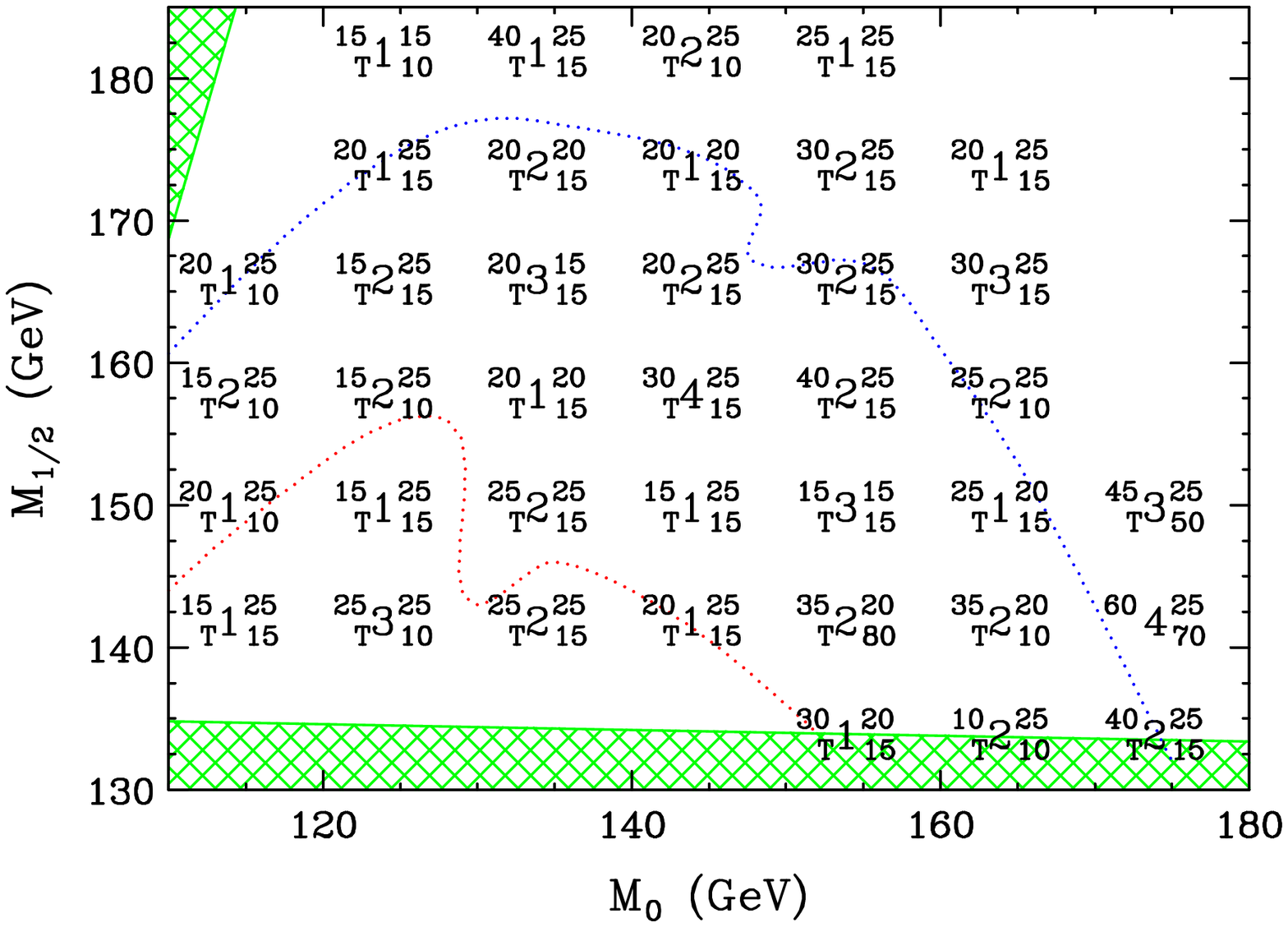,height=3.5in}}
\begin{center}
\parbox{5.5in}{
\caption[] {\small The same as Fig.~\ref{bc5_m0_small},
but for $\tan\beta=35$.
\label{bc35_m0_small}}}
\end{center}
\end{figure}

In Fig.~\ref{bc5_m0_small} (\ref{bc35_m0_small})
we show the optimum cuts chosen in our optimization procedure,
in the $M_0$, $M_{1/2}$ plane, for $\tan\beta=5$ ($\tan\beta=35$),
in the small $M_0$ region.
We use the following notation to describe the set of cuts at each
point. The central symbol indicates the set of lepton $p_T$ cuts: the
symbols ``1'' through ``5'' refer to
$\{11,5,5\},~\{11,7,5\},~\{11,7,7\},~ \{11,11,11\}$ and $\{20,15,10\}$
GeV lepton $p_T$ cuts, respectively.  The left superscript shows the
value (in GeV) of the low-end invariant mass cut
($m_{\ell^+\ell^-}^\gamma>10$ to 60 GeV). A left subscript ``T''
indicates that the cut on the transverse $\ell\nu$ mass was selected.
The right superscript shows the $\met$ cut: $\met>\{15,20,25\}$ GeV
(``15'',``20'',``25''), or no cut (no symbol).  A right subscript
denotes the high-end dilepton invariant mass cut:
$|m_{\ell^+\ell^-}-M_Z|>\{10,15\}$ GeV (``10'',``15'') or
$m_{\ell^+\ell^-}<\{50,60,70,80\}$ GeV
(``50'',``60'',``70'',``80''). And finally, a tilde over the central
symbol indicates that the luminosity limit came from requiring 5
signal events rather than $3\sigma$ exclusion.

In Fig.~\ref{bc5_m0_small} we see that in the regions where background
is an issue, the combination of the $m_T$ cut and a tighter low-end
dilepton mass cut $m^{\gamma}_{\ell^+\ell^-}\sim 20$ GeV is typically
preferred.  Indeed, we find that these additional cuts reduce the $WZ$
background by more than a factor of 3, from 4.1~fb (with soft cuts
\cite{BK}) to 1.2~fb. Notice, however, in the small $\tan\beta$ case
the transverse mass cut is never enough by itself, i.e. whenever it is
chosen, it is almost always supplemented with a
$m^{\gamma}_{\ell^+\ell^-}$ cut of 15 to 25 GeV (with the exception of
two points with high lepton $p_T$ cuts).  On the other hand, there are
significant regions where the low invariant mass cut
$m^{\gamma}_{\ell^+\ell^-}$ by itself is enough to kill the
background, and the transverse mass cut is not needed. In the large
$\tan\beta$ case the transverse mass is always chosen at small $M_0$,
but only occasionally at large $M_0$.

We should point out that the optimum cuts in Figs.~\ref{bc5_m0_small}
and \ref{bc35_m0_small} can be interpreted in two ways. First, for a
given total integrated luminosity, say 10 ${\rm fb}^{-1}$, one can
first roughly look up from Fig.~\ref{3l} the sensitivity reach of the
Tevatron. Then, for the parameter space well inside the sensitivity
region, the actual choice of cuts is not so crucial. However, as one
approaches the boundary of the sensitivity region, the choice of
optimum cuts as a function of parameter space (as opposed to a fixed,
non-optimized set of cuts) can enhance the reach by an additional
10-20 GeV along the $M_{1/2}$ direction~\cite{MP-3L}. Alternatively,
at a given parameter space point near the border, optimization can
reduce the total integrated luminosity required to observe or exclude
that point by up to a factor of two \cite{MP-3L}.

In conclusion, we find that the trilepton channel remains one of the
leading candidates for SUSY discovery at the Tevatron. The other two
channels are in a sense complementary, although not as powerful. The
dilepton plus tau jet channel can be combined straightforwardly with
the trilepton channel to maximally increase the reach. With the new
very important $W\gamma^*$ background included, the reach of course
suffers somewhat. We find that with only the standard Run 2 luminosity
of 2 fb$^{-1}$ the reach is quite limited. With the larger
background it is even more imperative that the Tevatron collect as
much luminosity as possible to have a decent chance at discovering
supersymmetry.

\section*{Acknowledgments}

We thank J.~Campbell, R.K.~Ellis, J.~Lykken, F.~Paige and
X.~Tata for useful discussions and H.~Baer for correspondence. 
K.T.M. (D.M.P.) is supported by Department of
Energy contract DE-AC02-76CH03000 (DE-AC02-98CH10886).

Note added: Our results shown in Figs.~\ref{3l}-\ref{2l1t} are similar
to those of Ref.~\cite{BDPQT} once the error in the original version
of \cite{BDPQT} was fixed \cite{XT}. We warn the reader that the
analysis of \cite{BDPQT} still neglects the $\gamma^\ast$
contribution and the interference effects in $ZZ$ production, and
lacks a detailed detector simulation for the $WZ$ process.

\end{document}